\documentclass{emulateapj}
\usepackage{apjfonts}
\usepackage{rotating}

\begin{document}

\title{KMT-2015-1\MakeLowercase{b}: a Giant Planet Orbiting a Low-mass Dwarf Host Star Discovered by a New High-cadence Microlensing Survey with a Global Telescope Network}

\author{
K.-H. Hwang$^{1}$,
C. Han$^{1,\ast}$,
J.-Y.Choi$^{1}$,
H. Park$^{1}$,
Y. K. Jung$^{1}$,
I.-G. Shin$^{1}$,
M. D. Albrow$^{2}$,
A. Gould$^{3}$,
V. Bozza$^{4,5}$,
\\
B.-G. Park$^{6}$,
S.-L. Kim$^{6}$,
C.-U. Lee$^{6}$,
S.-M. Cha$^{6,7}$,
D.-J. Kim$^{6}$
and
Y. Lee$^{6,7}$
}

\affil{$^{1}$Department of Physics, Chungbuk National University, Cheongju 361-763, Republic of Korea}
\affil{$^{2}$University of Canterbury, Department of Physics and Astronomy, Private Bag 4800, Christchurch 8020, New Zealand}
\affil{$^{3}$Department of Astronomy, Ohio State University, 140 W. 18th Ave., Columbus, OH 43210, USA}
\affil{$^{4}$Dipartimento di Fisica "E. R. Caianiello", Universit'a di Salerno, Via Giovanni Paolo II, 84084 Fisciano (SA), Italy}
\affil{$^{5}$Istituto Nazionale di Fisica Nucleare, Sezione di Napoli, Via Cintia, 80126 Napoli, Italy}
\affil{$^{6}$Korea Astronomy and Space Science Institute, Daejon 305-348, Republic of Korea}
\affil{$^{7}$School of Space Research, Kyung Hee University, Yongin 446-701, Republic of Korea}
\affil{$^{\ast}$Corresponding author}

\begin{abstract}
We report the discovery of an extrasolar planet, KMT-2015-1b,  
that was detected using the microlensing technique. The planetary 
lensing event was observed by KMTNet survey that has commenced in 
2015. With dense coverage by using network of globally distributed 
telescopes equipped with very wide-field cameras, the short planetary 
signal is clearly detected and precisely characterized.  We find that 
KMT-2015-1b is a giant planet orbiting a low-mass M-dwarf host 
star. The planet has a mass about twice that of Jupiter and it is 
located beyond the snow line of the host star.  With the improvement 
of existing surveys and the advent of new surveys, future microlensing 
planet samples will include planets not only in greatly increased number 
but also in a wide spectrum of hosts and planets, helping us to have a
better and comprehensive understanding about the formation and evolution 
of planets.
\end{abstract}

\keywords{gravitational lensing: micro -- planetary systems}

\section{INTRODUCTION}

Since the first discovery by \cite{wolszczan92} followed by \cite{mayor95},  
many exo-planets have been discovered. With the {\it Kepler} mission, 
the number of known planets explosively increased and now reaches 
$\sim$2000. Most of them were discovered by either the transit or 
radial-velocity methods.

Although relatively few, planets have also been discovered using the 
microlensing method.  Due to the fact that these planetary systems are 
detected through their gravitational fields rather than their radiation, 
this method makes it possible to detect planets around faint stars and 
even dark objects. Furthermore, microlensing is sensitive to planets on 
wide orbits beyond the snow line, which separates regions of rocky planet 
formation from regions of icy planet formation, while other major planet 
detection techniques are sensitive to close-in planets. Being able to 
detect planets that are difficult to be detected by other techniques, 
the method is important for the comprehensive understanding of planet
formation \citep{gaudi12}.

One of the important reasons for the small number of microlensing planets 
is the difficulty of observation. The planetary signal of a microlensing 
planet is a short-term perturbation to the smooth lensing light curve 
induced by the host star \citep{mao91, gould92b}.  To detect short planetary 
signals, the previous generation of lensing experiments applied a strategy 
where lensing events were detected by wide-field surveys and events detected 
by surveys were intensively monitored using multiple narrow-field telescopes 
\citep{albrow98}.  However, the limited number of follow-up telescopes makes 
it difficult to intensively monitor all alerted events, which number reaches 
several hundreds at a given moment. As a result, the detection efficiency of 
microlensing planets has been low.

In this paper, we report the discovery of a giant planet KMT-2015-1b 
orbiting a low-mass M dwarf.  The planetary lensing event was observed by a 
new survey that achieves round-the-clock coverage of lensing events with a 
high cadence by using global network of telescopes equipped with very 
wide-field cameras.

The paper is organized as follows. In Section 2, we describe 
the observation of the planetary microlensing event by the new 
lensing survey, including instrument, data acquisition and 
reduction process. In Section 3, we give a description of the 
modeling procedure conducted to analyze the observed lensing 
light curve. We provide the estimated physical parameters of
 the discovered planetary system in Section 4. Finally, we summarize 
the result and make a brief discussion about the result in Section 5.

\begin{figure*}[th]
\epsscale{0.85}
\plotone{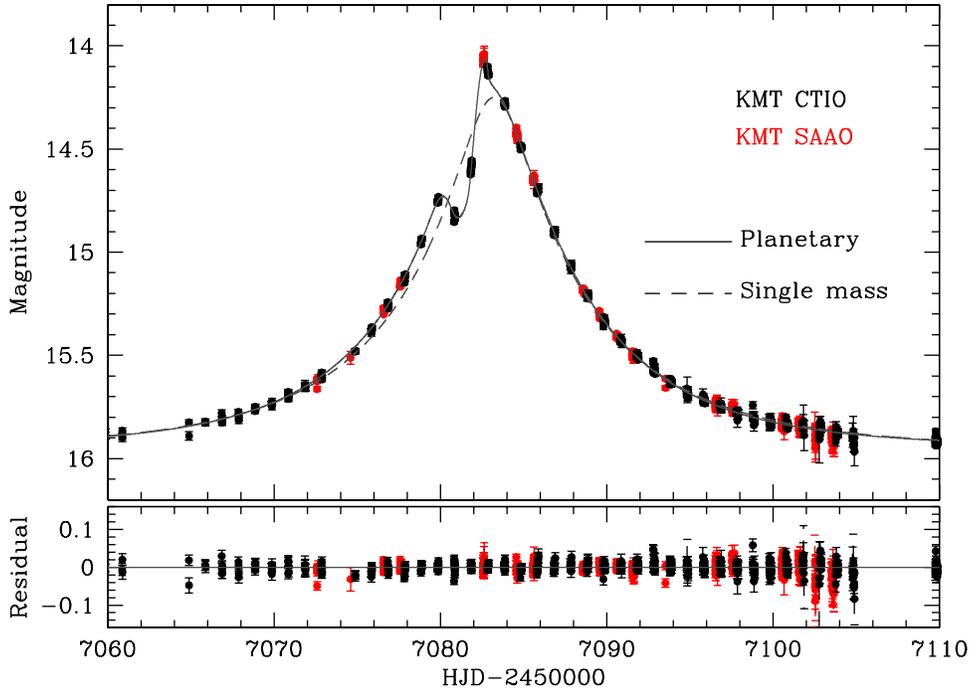}
\caption{\label{fig:one}
Light curve of the microlensing event KMT-2015-BLG-0048.
Solid and dashed curves represent the best-fit models from
binary (planetary) and single-mass lens modeling, respectively.
}
\end{figure*}

\section{OBSERVATION}

The planet was discovered from the observation of
a microlensing event that occurred on a star located toward
the Galactic bulge field. The equatorial coordinates of the lensed star (source) are
(RA,DEC) = 
($17^{\rm h}58^{\rm m}39^{\rm s}\hskip-2pt.01$,$-28\arcdeg01\arcmin54\arcsec\hskip-2pt.1$),
that correspond to the Galactic coordinates
($l$,$b$) = ($2.24\arcdeg$,$-2.00\arcdeg$).
The lensing-induced brightening of the source star was noticed in
early February, 2015 
by the  OGLE \citep{udalski15} survey
and lasted for about a month, during which the bulge field can be
seen for less than an hour per night.

The event was also observed by the KMTNet (Korea Microlensing Telescope Network) 
lensing survey that started its test observation in February, 2015, 
which matches the occurrence time of the event. 
The event was dubbed as KMT-2015-BLG-0048 in the KMT event list.
The survey uses three identical telescopes that are located at Cerro
Tololo Interamerican Observatory, Chile (KMT CTIO), South African Astronomical
Observatory, South Africa (KMT SAAO), and Siding Spring Observatory, Australia (KMT
SSO). At the time of the event, KMT SSO was not online and the event was observed
by two telescopes, KMT CTIO and KMT SAAO. Each telescope has a 1.6m aperture and
is equipped with a mosaic camera composed of four 9K$\times$9K CCDs. Each CCD has a pixel
size of 10 microns corresponding to 0.4 arcsec/pixel and thus the camera has 4 ${\rm deg}^2$
field of view \citep{kim15}. With the global network of wide-field telescopes that are separated by $\sim$8
hrs in longitudes, the survey continuously observes the Galactic bulge field with a
cadence of 1 observation every 10 minutes. Images were taken in Johnson-Cousin $I$ and
$V$ bands. From the observation of the event, we acquired 786 $I$-band and 54 $V$-band
images from KMT CTIO and 1117 $I$-band images from KMT SAAO observations.
We note that $I$-band data are used to analyze the lensing event and 
$V$-band data are used to constrain the source star.

Photometry of the images were conducted using a customized pipeline that is based on
the difference image analysis code of \cite{albrow09}
Since data are taken by the different telescopes
and photometry based on an automated pipeline often results in inaccurate error
estimation, we renormalize error bars of the individual data sets by
\begin{equation}
\sigma ' = k ( {\sigma_0^2} + {\sigma_{\rm min}^2} ) ^{1/2} ,
\end{equation}
where $\sigma_0$ is the error estimated from the pipeline, 
$\sigma_{\rm min}$ is a factor used to make the cumulative distribution 
function of $\chi^2$ as a function of lensing magnification linear, 
and $k$ is a scaling factor to make $\chi^2$ per degree of freedom 
(dof) become unity. We note that the factor $\sigma_{\rm min}$ is required
to ensure that each data set is fairly weighted according to its error bars.

In Figure~\ref{fig:one}, we present the light curve of KMT-2015-BLG-0048. 
Compared to the continuous and symmetric light curve of a single-mass event, 
the light curve exhibits a short-term perturbation during 
$7080.0\lesssim{\rm HJD}-2450000\lesssim7082.5$. The perturbation shows a 
feature that is composed of a depression centered at ${\rm HJD}-2450000\sim7081.5$ 
and brief bumps at both edges of the depression.  Such dips, usually surrounded 
by two bumps, are a generic feature of lensing systems with small mass ratios 
$q\ll1$ with normalized planet-star separations $s<1$, i.e., planets inside 
the Einstein ring.  This is because the star's gravity generates two images, 
one inside and the other outside the Einstein ring.  The former, being a 
saddle point on the time delay surface, is easily suppressed if the planet 
lies in or near the path, thereby causing relative demagnification, and 
hence a dip in the light curve.  To be noted is that the major structure 
of the anomaly feature was well covered by the survey observation despite 
the short time window each night toward the field.

\begin{deluxetable}{lc}
\tablecaption{Lensing parameters\label{table:one}}
\tablewidth{0pt}
\tablehead{
\multicolumn{1}{c}{Parameters}     &
\multicolumn{1}{c}{Values} 
}
\startdata
$\chi^2/{\rm dof}$         &    1896/1896  \\     
$t_0 \ ({\rm HJD})$        & 2457083.081 $\pm$ 0.004 \\     
$u_0$                      &       0.221 $\pm$ 0.003 \\     
$t_{\rm E} \ ({\rm days})$ &      10.90  $\pm$ 0.10  \\     
$s$                        &       0.955 $\pm$ 0.003 \\     
$q \ (10^{-3})$            &       6.84  $\pm$ 0.18  \\     
$\alpha \ ({\rm rad})$     &       5.352 $\pm$ 0.003 \\     
$\rho \ (10^{-3})$         &      44.7   $\pm$ 0.8
\enddata
\end{deluxetable}

\begin{figure}[th]
\epsscale{1.15}
\plotone{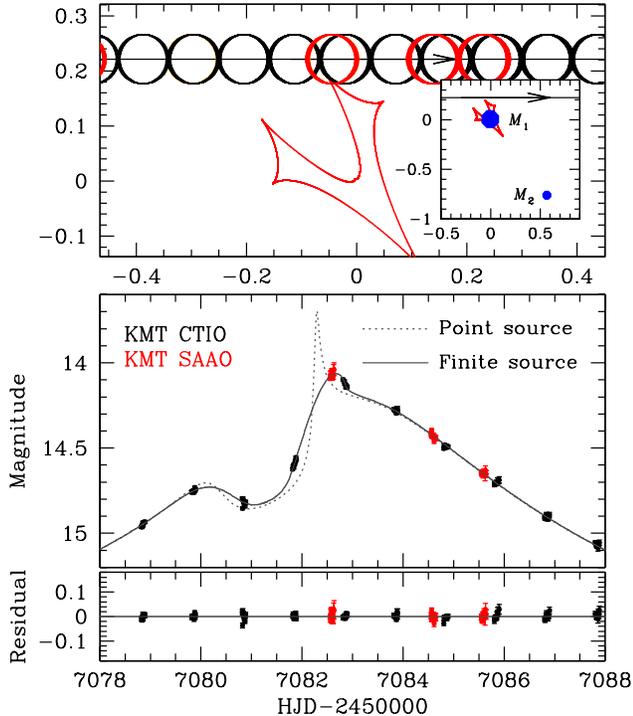}
\caption{\label{fig:two}
Lens system geometry. The upper panel shows the source
trajectory (straight line with an arrow) with respect to the lens
components (marked by $M_1$ and $M_2$) and caustics (closed concave curve) 
and the lower panel shows the light variation with the
progress of the source position. Lengths are scaled to the Einstein radius
and the source trajectory is aligned so that the progress of the source
matches the light curve shown in the lower panel. The inset in the upper
panel shows the wide view and the major panel shows the enlarged view
around the caustic. The empty circles on the source trajectory represent
the source positions at the times of observation and the size indicates the
source size. The dotted curve in the lower panel is the light curve
expected for a point source.
}
\end{figure}

\section{ANALYSIS}
Keeping in mind that the anomaly pattern is likely to be 
produced by a binary lens with a low mass ratio, we conduct 
binary-lens modeling. Basic binary-lens modeling requires
at least 9 parameters, including 7 principal parameters 
for the lensing system and 2 flux parameters 
for each observatory. The first 3 of the principal parameters
describe the source approach with respect the lens, 
including the time of the closest source approach to the 
reference position of the lens, $t_0$, the lens-source separation at
that moment, $u_0$ (impact parameter), and the time for the source 
to cross the angular Einstein radius $\theta_{\rm E}$ of the lens, 
$t_{\rm E}$ (Einstein time scale). For the reference position of the
lens, we use the center of mass. The other 3 principal 
parameters describe the lens binarity including the projected 
separation $s$ and the mass ratio $q$ between the binary
components, and the angle between the source trajectory and 
the binary axis, $\alpha$. We note that the impact parameter 
$u_0$ and the binary separation $s$ are normalized by $\theta_{\rm E}$.
The other parameter defined as the ratio of the angular source radius 
to the Einstein radius, $\rho = \theta_\ast / \theta_{\rm E}$, 
is needed to describe light curve deviations affected by 
finite-source effects.

\begin{figure}[th]
\epsscale{1.15}
\plotone{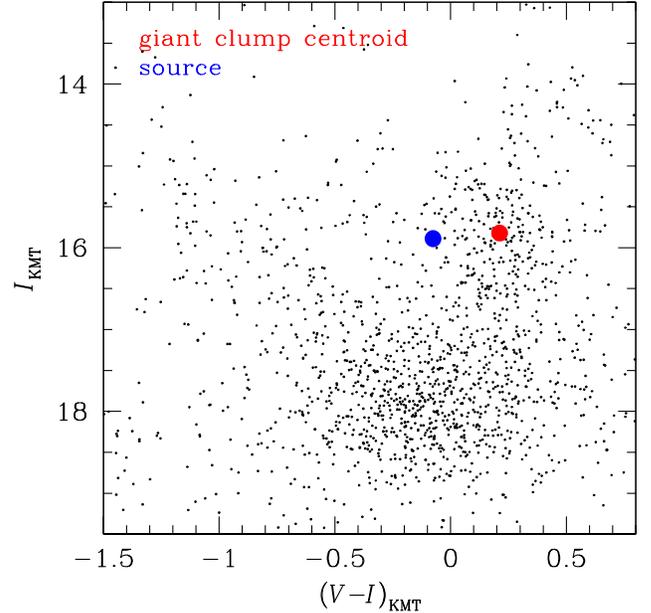}
\caption{\label{fig:three}
Source position in the instrumental color-magnitude diagram 
of nearby stars with respect to the centroid of the giant clump.
}
\end{figure}

For some lensing events, the basic parameters are not adequate 
to precisely describe observed lensing light curves. 
The known causes of such deviations include the parallax effect
\citep{gould92a} and the lens orbital effect 
\citep{albrow00, an02, jung13}. The parallax effect is
caused by the deviation of the observer's motion from 
rectilinear due to the orbital motion of the Earth around the Sun. 
On the other hand, the lens-orbital effect is caused
by the orbital motion of the lens. Such effects are important 
for long time-scale events for which the orbital motion of 
the Earth and/or lenses are important. For KMT-2015-BLG-0048, 
which lasted for $\sim$1 month, we find that these 
higher-order effects are negligible.

Modeling the light curve proceeded in several steps. 
First, we conduct thorough a grid search for solutions in 
the parameter space of ($s$, $q$, $\alpha$), 
for which lensing light curves vary sensitively to the change 
of the parameters. In this process, other parameters are
searched for by using a downhill approach. Second, we investigate 
all possible local solutions in the parameter space in order 
to check the existence of degenerate solutions where different 
combinations of the lensing parameters result in a similar light curve.
Finally, we search for the global solution by comparing 
$\chi^2$ values of the identified local solutions.
For the downhill $\chi^2$ minimization, we use the
Markov Chain Monte Carlo (MCMC) method.

From the search for a solution, we find that the event
 was produced by a planetary system where the planet/host 
mass ratio is $q = 6.84\times10^{-3}$ and the projected planet-host
separation, $s=0.96$, is slightly less than 
the Einstein radius of the lens. 
In Table~\ref{table:one}, we present the best-fit 
lensing parameters, where the error bars are estimated 
from the scatter of the MCMC chain.
In Figure~\ref{fig:two}, we present the geometry of the lens
system corresponding to the solution. Due to the resonance of the 
projected separation to $\theta_{\rm E}$, i.e.\ $s\sim1$, 
the lens system forms a single large caustic around the host of the
planet. The source passed the backside of the arrowhead-shaped caustic.
The depression in the light curve occurred when the source was 
in the demagnification valley between the two protrudent cusps that caused 
the brief bumps on both sides of the depression. The source crossed 
the tip of one of the cusps during which the light curve shows a clear
finite-source signature from which we accurately measure the normalized 
source radius $\rho$. To be noted is that the measured value of 
$\rho = 44.7\times10^{-3}$ is significantly larger than typical values 
$\sim$10$^{-3}$ for main-sequence and $\sim$10$^{-2}$ for giant 
source stars. For a given size of a bulge star, this suggests 
that the Einstein radius is very small.

Since finite-source effects are clearly detected, it is possible to 
determine the angular Einstein radius. The Einstein radius is determined by
$\theta_{\rm E} = {\theta_\ast / \rho}$.
The normalized source radius $\rho$ is measured from the analysis 
of the light curve around the planetary perturbation. 
The angular radius of the source star, $\theta_\ast$, 
is determined from its de-reddened color $(V-I)_0$ and brightness $I_0$. 
To measure color and brightness, we locate the source star in 
the instrumental(uncalibrated) color-magnitude diagram of neighboring 
stars in the same field and calibrate the color and 
brightness based on the location of the centroid 
of the giant clump (GC), for which its de-reddened color $(V-I)_{0, \rm GC}$ 
and brightness $I_{0, \rm GC}$ are known to be constant and thus can 
be used as a standard candle. Figure~\ref{fig:three} shows the locations of
the source and giant clump centroid in the instrumental color-magnitude diagram. 
By adopting
$(V-I)_{0, \rm GC}=1.06$ \citep{bensby11} and
brightness $I_{0, \rm GC}$ accounting for variation with Galactic longitude 
\citep{nataf13}, we find that 
$(V-I, I)_0 = (0.78, 14.5)$,
indicating that the source is a G-type giant star. 
We then convert $V-I$ into $V-K$ using the color-color relation of 
\cite{bessel88} and finally determine $\theta_\ast$
using the color-angular radius relation of 
\cite{kervella04}. 
We find that the angular source radius is 
$\theta_\ast=4.39\pm0.56\ {\rm{\mu}as}$. 
Then the angular Einstein radius is
\begin{equation}
\theta_{\rm E}=0.098\pm0.013\ {\rm{mas}} .
\end{equation}
Combined with the Einstein time scale estimated from lens modeling,
the relative lens-source proper motion is determined as
\begin{equation}
\mu={\theta_{\rm E} \over t_{\rm E}}=3.29\pm0.43{\ }{\rm{mas}{\ }yr^{-1}} .
\end{equation}
As expected, the estimated Einstein radius is substantially smaller than 
$\sim0.5$ mas for typical Galactic microlensing events.

\section{PHYSICAL PARAMETERS}

For the unique determination of the mass $M$ and distance $D_L$ to the 
lens, it is required to simultaneously measure both the lens parallax 
and the Einstein radius \citep{gould92a}. Nevertheless, the above 
measurements of the Einstein radius $\theta_{\rm E}$ and the proper 
motion $\mu$ can be combined with a Galactic model to statistically 
constrain $M$ and $D_L$ via a Bayesian analysis.  We use the \citet{han95} 
model, whose matter distribution is based on a double-exponential disk and 
a triaxial bulge.  The disk velocity distribution is assumed to be Gaussian 
about the rotation velocity and the bulge velocity distribution is a 
triaxial Gaussian with components deduced from the flattening via the 
tensor virial theorem.  The mass function is based on 
{\it Hubble Space Telescope} star counts.

The results of this Bayesian analysis 
\begin{equation}
M=0.18 \pm 0.12\ M_\odot;\qquad D_L=8.2\pm0.9 \ {\rm kpc}
\end{equation}
can be  easily understood based on simple physical reasoning. First, 
the Enstein radius is by definition the root-mean-square to the lens mass 
and the lens-source relative parallax $\pi_{\rm rel}$, i.e.
\begin{equation}
\theta_{\rm E}=\sqrt{\kappa M \pi_{\rm rel}};\qquad
\kappa={4G\over c^2 {\rm AU}} \simeq 8.1 {{\rm mas}\over M_\odot}.
\end{equation}
Because $\theta_{\rm E}=0.098\ {\rm mas}$ is unusually small, either $M$ or 
$\pi_{\rm rel}$ should be small:
$(M/M_\odot)(\pi_{\rm rel}/\mu{\rm as})=1.2$.
Thus, if the lens is a star ($M > 0.08\ M_\odot$), it is $\pi_{\rm rel}<15\ \mu{\rm as}$
and so lies in the Galactic bulge.  This, combined with the fact that 
the Galactic density profile along the line of sight peaks strongly in 
the bulge, heavily favors bulge lenses.  Finally, the measured proper 
motion is very typical of bulge lenses, but only half the size of typical 
proper motions.

In principle, the lens could be a star of any mass in the bulge. However, 
the measurement of $\theta_{\rm E}$ telles us that if it were a solar mass star, 
then $\pi_{\rm rel}=1.2\ \mu{\rm as}$, which would imply a distance from lens 
to source of only $D_S-D_L=70$ pc. It is this small amount of available phase 
space for heavier lenses that drive the Bayesian estimate toward low mass lenses.

The measurement of $q$ from the lens model then directly yields an estimate 
of the planet mass
\begin{equation}
M_p=qM=2.2 \pm 1.4\ M_J,
\end{equation}
while the measurement of $s$ yields the projected lens-host separation
\begin{equation}
d_\perp = s D_L \theta_{\rm E} = 0.76 \pm 0.08 \ {\rm AU}.
\end{equation}
Considering the low temperature of the host star, the planet is located 
beyond the ice line.

\section{SUMMARY AND DISCUSSION}
We reported the discovery of an extrasolar planet
KMT-2015-1b that was detected by using
the microlensing technique. The planetary lensing
event was observed by the KMTNet survey that has commenced in 2015.
Despite the short time window in the early bulge season, 
unambiguous detection and precise characterization of the planetary system 
was possible due to the dense coverage of the planet-induced perturbation 
by using network of globally distributed telescopes equipped with 
very wide-field cameras. We find that KMT-2015-1b is a giant planet 
orbiting a low-mass M-dwarf host star. The planet has a mass about 
twice that of the Jupiter and it is located beyond the snow line of the host star.

Cool M dwarfs far outnumber sun-like stars and thus understanding 
the process of planet formation around them is important. Furthermore, 
small masses and low luminosities of M dwarfs provide leverage
on conditions of planet formation, enabling to check the validity 
of existing formation theories and refine survived theories. 
With the improvement of existing surveys and the advent of new surveys, 
future microlensing planet sample will include planets not only 
in greatly increased number but also in wide spectrum of hosts and planets, 
helping us to have better and comprehensive understanding 
about the formation and evolution of planets.

\acknowledgments
Work by C.H. was supported by Creative Research Initiative
Program (2009-0081561) of National Research Foundation of Korea.
Work by A.G. was supported by NASA grant NNXRAB99G.
This research has made the telescopes of KMTNet operated by the
Korea Astronomy and Space Science Institute (KASI).
We acknowledge the high-speed internet service (KREONET)
provided by Korea Institute of Science and Technology Information (KISTI)

\end{document}